\documentclass[conference]{IEEEtran}
\IEEEoverridecommandlockouts

\usepackage{amssymb}
\usepackage{amsfonts}
\usepackage{amsmath}

\usepackage{algorithm}
\usepackage{algorithmic}
\usepackage{acronym}
\usepackage{textcomp}
\usepackage{xcolor}
\usepackage{graphicx}
\usepackage{cite}
\usepackage{comment}
\usepackage{balance}
\usepackage{acronym}
\usepackage{tabularx}
\usepackage{booktabs}
\usepackage{enumitem,bbm}
\usepackage{wrapfig}
\usepackage{hyperref}
\usepackage{soul}

\usepackage{IZ_Shortcuts}
\usepackage[section]{placeins}
\usepackage{float}
\usepackage{adjustbox}

\usepackage{tikz}
\usetikzlibrary{arrows.meta,positioning,calc,shadows.blur,decorations.pathreplacing}
\usetikzlibrary{positioning,calc,decorations.pathreplacing,shadows.blur}
\definecolor{leftcol}{RGB}{255,202,128}
\definecolor{rightcol}{RGB}{173,216,230}
\definecolor{stftcol}{RGB}{210,210,255}
\definecolor{istftcol}{RGB}{210,255,210}
\definecolor{specbg}{RGB}{245,245,245}

\def\BibTeX{{\rm B\kern-.05em{\sc i\kern-.025em b}\kern-.08em
T\kern-.1667em\lower.7ex\hbox{E}\kern-.125emX}}

\title{Interpretable Binaural Deep Beamforming Guided by Time-Varying Relative Transfer Function \thanks{This work was supported by the Israel Science Foundation (ISF) and the German Research Foundation (DFG) through the ISF–DFG Joint Research Program, Grant No.~1280/25.}}

\author{
\IEEEauthorblockN{Ilai Zaidel}
\IEEEauthorblockA{\textit{Bar-Ilan University}, Israel\\
ilai.zaidel@biu.ac.il; 0009-0005-4578-6742}
\and
\IEEEauthorblockN{Sharon Gannot}
\IEEEauthorblockA{\textit{Bar-Ilan University}, Israel\\
sharon.gannot@biu.ac.il; 0000-0002-2885-170X}
}

\begin{document}
\maketitle
\acrodef{RTF}{relative transfer function}
\acrodef{RTFs}{relative transfer functions}
\acrodef{SI-SDR}{scale-invariant signal-to-distortion ratio}
\acrodef{DOA}{direction of arrival}
\begin{abstract}
In this work, we propose a deep beamforming framework for speech enhancement in dynamic acoustic environments. The framework learns time-varying beamformer weights from noisy multichannel signals via a deep neural network, guided by a continuously tracked relative transfer function (RTF) of a moving target speaker. We analyze the network’s spatial behavior on an 8-microphone linear array by evaluating narrowband and wideband beampatterns in three modes: (i) oracle guidance with true RTFs, (ii) guidance with subspace-tracked RTF estimates, and (iii) operation without RTF guidance. Results show that RTF guidance yields smoother, more spatially consistent beampatterns that track the target direction of arrival (DOA), whereas the unguided model fails to maintain a clear spatial focus. We further extend the framework to binaural beamforming for dynamic target-speaker enhancement. The system is trained using a head-related transfer function (HRTF)-based acoustic simulation of a moving source, enabling realistic spatial rendering at the left and right ears. Spatial cue preservation is quantitatively evaluated in terms of interaural level differences (ILD) and interaural time differences (ITD), demonstrating the method’s suitability for hearable applications.


\end{abstract}


\section{Introduction}
\label{sec:intro}

Speech enhancement algorithms aim to improve the perceptual quality and intelligibility of noisy speech signals in acoustic environments. In multichannel setups, spatial diversity can be exploited to reduce noise or to separate the target speaker. In this work, we address only the noise-reduction task, namely, enhancing a single desired speech signal contaminated by noise.
Classical beamforming approaches, such as the minimum variance distortionless response (MVDR) beamformer \cite{Gannot:Consolidated_Prespective}, have demonstrated strong performance under static acoustic conditions. Several studies have shown that employing the relative transfer function (RTF) as the steering vector in an MVDR beamformer leads to improved speech quality compared to conventional formulations that rely solely on the direct-path acoustic propagation \cite{gannot2001RTF,Shmaryahu_Gannot_acoustic_reflections_2022}. However, the effectiveness of such approaches critically depends on accurate estimation of the steering vector. In dynamic scenarios, where the acoustic transfer functions (ATFs) and their corresponding RTFs evolve over time, reliable tracking of these time-varying RTFs becomes essential.

Recently, DNN-based beamformers have achieved strong performance by jointly learning spatial and spectral representations from data \cite{FasNET,Ren:Causal_UNet_Beamforming}. By replacing analytically derived weights with trainable networks, they capture nonlinear mappings and adapt to diverse acoustic conditions. However, their spatial behavior is often non-interpretable, motivating methods that explicitly analyze the spatial properties of the multichannel algorithms and encourage their spatial selectivity  \cite{Kellermann_Spatial__info_TSE,tesch2022insights,Cohen:ExNet_BF_PF}.
The central role of the RTF in classical beamforming motivates its integration into DNN-based beamformers. Whereas prior work has estimated RTF-based classical beamformer coefficients \cite{Gannot:Consolidated_Prespective,ronai2025rtf,RTF_estimation_correlations} or used it as a spatial filter within deep architectures \cite{Align_and_Filter}, our approach feeds the RTF as an input feature into the network.
%
%
Recent results in \cite{eisenberg2025end} for target speaker extraction (TSE) show that RTF-based spatial guidance outperforms DOA-based guidance. This likely stems from the RTF’s ability to capture complex acoustic propagation in reverberant environments more faithfully \cite{Shmaryahu_Gannot_acoustic_reflections_2022}.

Dynamic scenarios are inherently more challenging than static settings, especially when spatial interpretability is required. In \cite{timo_self_steering}, a spatially selective deep filter was introduced that depends only on weak spatial guidance derived from the target speaker’s initial position.


In binaural speech enhancement, preserving spatial cues is crucial, and cue preservation is typically assessed via interaural measures such as ILD/ITD \cite{Doclo2024Binaural}.
Prior work has shown that deep binaural speech separation can handle moving speakers while preserving spatial cues \cite{han2021binaural,han2023online}. More recently, binaural TSE methods have been proposed that explicitly leverage HRTFs as spatial cues, enabling accurate cue preservation in both anechoic and reverberant environments \cite{ellinson2025binaural}.

In this paper, we propose guiding a DNN-based beamformer using time-varying RTF estimates of a moving target speaker, tracked with the projection approximation subspace tracking (PAST) algorithm \cite{Yang:PAST}. This blind tracking approach has been applied to speech enhancement and extended to dynamic multichannel settings \cite{Affes_Grenier_subspace_tracking,golan2010subspace}, making it a natural candidate for providing spatial guidance. Building on the explainable beamforming architecture \cite{Cohen:ExNet_BF_PF}, we incorporate the tracked RTF as an additional input feature and demonstrate improved spatial performance compared with the same architecture without RTF guidance. Finally, we extend the framework to a binaural configuration and show that time-varying RTF guidance also improves spatial cue preservation (ILD/ITD).


\section{Problem Formulation}
\label{sec:problem_formulation}

In the short-time Fourier transform (STFT) domain, the multichannel mixture signal is modeled as
\begin{equation}\label{eq:stft_model}
\vct{y}(l,k)=\vct{h}(l,k)s(l,k)+\vct{n}(l,k),
\end{equation}
where $l$ and $k$ denote the time-frame and frequency-bin indices, respectively. Here, $s(l,k)$ represents the target speech component, $\vct{h}(l,k)$ contains the acoustic transfer functions (ATFs) from the source to the microphones, and $\vct{n}(l,k)$ denotes additive noise. We consider a dynamic scenario in which the ATFs are time-varying.

We apply two time-varying spatial filters:
\begin{subequations}\label{eq:bf}
\begin{gather}
\hat{s}_{\text{L}}(l,k)=\mathbf{w}^{\mathrm{H}}_{\text{L}}(l,k)\vct{y}(l,k), \label{eq:bf_L}\\
\hat{s}_{\text{R}}(l,k)=\mathbf{w}^{\mathrm{H}}_{\text{R}}(l,k)\vct{y}(l,k), \label{eq:bf_R}
\end{gather}
\end{subequations}
where $\hat{s}_{\text{L}}(l,k)$, $\hat{s}_{\text{R}}(l,k)$ are the binaural outputs and $\mathbf{w}_{\text{L}}(l,k),\mathbf{w}_{\text{R}}(l,k)$ are the DNN-based  beamformer weights.

Our goal in this work is twofold: (i) to estimate the clean speech signal while preserving the target’s spatial cues, and (ii) to achieve interpretable spatial filtering, characterized in terms of beampatterns. Both objectives are addressed under dynamic acoustic conditions.

Binaural cue preservation (ILD/ITD) is evaluated in a configuration with a single microphone per ear. In this case, $\vct{h}(l,k)$ corresponds to the listener’s binaural HRTFs.
The spatial filtering characteristics are further examined using an 8-microphone array. Since no HRTFs are incorporated in this simulation, binaural rendering is not modeled. Therefore,  spatial behavior is evaluated purely through beampattern analysis of \eqref{eq:bf_L}. Similar analysis can be applied to \eqref{eq:bf_R}.

In both steups, we assume that the reverberation of the target speaker is negligible due to the close proximity between the speaker and the listener.

\section{Proposed Method}
\label{sec:proposed_method}

This section details the network architecture, which comprises two parallel branches: one estimates the left-ear signal and the other the right-ear signal; together they constitute the binaural output.


The proposed U-Net design follows \cite{Cohen:ExNet_BF_PF}. Two identical U-Nets estimate the complex filter-and-sum beamformer weights, $\mathbf{w}_{\mathrm{L}}(l,k)$ and $\mathbf{w}_{\mathrm{R}}(l,k)$. Both operate on the same multichannel STFT input $\mathbf{y}(l,k)$ but differ in training targets: the left branch reconstructs a clean left-ear reference signal, and the right branch a clean right-ear reference. Each branch is guided by a time-varying RTF estimate, $\{\hat{\mathbf{a}}_{\mathrm{L}},\hat{\mathbf{a}}_{\mathrm{R}}\} \in \mathbb{C}^{M \times F \times L}$, with $M$ microphones, $F$ frequency bins, and $L$ time frames. The RTFs are obtained by normalizing the ATF using different reference microphones, the left-most for $\hat{\mathbf{a}}_{\mathrm{L}}$, and right-most for $\hat{\mathbf{a}}_{\mathrm{R}}$.\footnote{In the two-microphone case, the reference microphones correspond to the left- and right-ear microphones.} The full architecture is shown in Fig.~\ref{fig:proposed model}. The various blocks are now detailed.

\subsection{U-Net Model with Attention Fusion}


We integrate an attention-based fusion frontend prior to the U-Net to align the RTF with the multichannel noisy input. The U-Net adopts an encoder-decoder with skip connections, following \cite{Cohen:ExNet_BF_PF,Ren:Causal_UNet_Beamforming} with task-specific modifications: eight convolutional blocks (batch normalization, dropout, LeakyReLU) in the encoder and transposed-convolution blocks in the decoder \cite{Cohen:ExNet_BF_PF}. Attention gates are integrated into the skip connections to emphasize relevant encoder features \cite{Cohen:ExNet_BF_PF}. See \cite{Cohen:ExNet_BF_PF} (Fig.~2) for the baseline U-Net model.



\begin{figure*}[t]
\vspace*{-4mm} 

\centering
\begin{adjustbox}{max width=\textwidth}
\tikzset{
  >=Latex,
  block/.style={draw, rounded corners=3pt, thick, align=center,
                minimum width=20mm, minimum height=10mm, blur shadow},
  blockISTFT/.style={draw, rounded corners=3pt, thick, align=center,
                minimum width=20mm, minimum height=10mm},                
  tinyblock/.style={draw, rounded corners=2pt, thick, align=center,
                    minimum width=12mm, minimum height=8mm, fill=specbg},
  tinyblockATTN/.style={draw, rounded corners=2pt, thick, align=center,
                    minimum width=12mm, minimum height=8mm, fill=specbg, blur shadow},
  arrow/.style={-Latex, very thick},
  stream/.style={draw, thick, minimum width=8mm, minimum height=18mm,
                 rounded corners=2pt, fill=white},
  mult/.style={
    draw, circle, thick, fill=white, inner sep=0pt,
    minimum size=6mm,     
    path picture={
      \draw[line width=0.9pt]
        ($(path picture bounding box.south west)+(1.0pt,1.0pt)$) --
        ($(path picture bounding box.north east)+(-1.0pt,-1.0pt)$);
      \draw[line width=0.9pt]
        ($(path picture bounding box.north west)+(1.0pt,-1.0pt)$) --
        ($(path picture bounding box.south east)+(-1.0pt,1.0pt)$);
    },
  },
waveglyph/.style={},
  pic wave/.style n args={3}{ 
    code={
      \def\W{#1}
      \def\H{#2}
      \def\C{#3}
      \draw[line width=0.6pt]
        plot[domain=0:\W, samples=60]
        (\x,{ \H*sin(360*\C*\x/\W) });
    }
  },
  wavstack/.style={
    thick, rounded corners=2pt, fill=white,
    minimum width=8mm, minimum height=18mm
  },
  pics/wave/.style n args={3}{
  code={
    \def\W{#1}
    \def\H{#2}
    \def\C{#3}
\def\amp{
  (
    \H * (
         0.25
       + 0.85*exp(-18*((\x+0.5)-0.25)^2)
       + 0.65*exp(-22*((\x+0.5)-0.55)^2)
       + 0.45*exp(-26*((\x+0.5)-0.82)^2)
    )
    * abs(
        0.90*sin(360*\C*(10*(\x+0.5)))
      + 0.35*sin(360*\C*(20*(\x+0.5)) + 20)
      + 0.18*sin(360*\C*(30*(\x+0.5)) - 55)
    )
  )
}

\draw[line width=0.9pt, line cap=round]
  plot[domain=-0.5:0.5, samples=19, variable=\x, ycomb]
  ({\x*\W},{\amp});

\draw[line width=0.9pt, line cap=round]
  plot[domain=-0.5:0.5, samples=19, variable=\x, ycomb]
  ({\x*\W},{-\amp});
  }
},
}
\newcommand{\bigvdots}{%
\begin{tikzpicture}[baseline=-0.6ex]
  \fill (0,0) circle (1.2pt);
  \fill (0,-0.18) circle (1.2pt);
  \fill (0,-0.36) circle (1.2pt);
\end{tikzpicture}%
}

\begin{tikzpicture}[
  font=\small,
  every node/.style={font=\small}
]
\node[
  wavstack,
  label={[align=center, xshift=-3mm, yshift=2mm]above:
    \large Multichannel noisy\\ \large inputs}
] (in) {};
\pic at ($(in.north)+(-3mm,-2mm)$) {wave={14.5mm}{3.2mm}{1.2}};

\node at ($(in.center)+(-3mm,0mm)$) {\bigvdots};

\pic at ($(in.south)+(-2.5mm,2mm)$) {wave={14.5mm}{3.2mm}{1.2}};
\node[block, fill=stftcol, right=10mm of in] (stft) {\Large STFT};

\coordinate (split) at ($(stft.east)+(14mm,0)$);

\draw[arrow] (in) -- (stft);
\draw[arrow] (stft.east) -- (split);
\draw[arrow] (stft.east) -- node[above,yshift=1pt] {\LARGE $\mathbf y$} (split);

\node[block, fill=leftcol!60, right=22mm of stft, yshift=10mm] (rtfL) {\large RTF\\ \large Estimation};

\node[tinyblockATTN, fill=leftcol!45, right=10mm of rtfL] (attnL) {\large Attn.};

\node[block, fill=leftcol!80, right=10mm of attnL] (unetL){\large U\text{-}Net};

\node[mult, right=15mm of unetL] (mulL) { };
\node[tinyblock, right=10mm of mulL] (specL) {\LARGE $\hat{s}_{\mathrm{L}}$};
\node[blockISTFT, fill=istftcol, right=10mm of specL] (istftL) {\large ISTFT};

\coordinate (outX) at ($(istftL.east)+(18mm,0)$);

\node[draw, rounded corners=2pt, minimum width=16mm, minimum height=10mm, align=center]
  (outL) at (outX |- istftL.center) {\large Left\\ \large out};

\draw[arrow] (split) |- (rtfL.west);          
\draw[arrow] (rtfL) -- (attnL);               
\draw[arrow] (attnL) -- (unetL);              


\draw[arrow] (split) |- ++(0,24mm) -| (attnL.north);

\draw[arrow] (unetL) -- node[above] {\LARGE $\mathbf w_{\mathrm L}$} (mulL);

\draw[arrow] (split) |- ++(0,24mm) -| (mulL.north);

\draw[arrow] (mulL) -- (specL);
\draw[arrow] (specL) -- (istftL);
\draw[arrow] (istftL) -- (outL);

\node[block, fill=rightcol!60, right=22mm of stft, yshift=-10mm] (rtfR)  {\large RTF\\ \large Estimation};

\node[tinyblockATTN, fill=rightcol!45, right=10mm of rtfR] (attnR) {\large Attn.};

\node[block, fill=rightcol!80, right=10mm of attnR] (unetR){\large U\text{-}Net};

\node[mult, right=15mm of unetR] (mulR) {};
\node[tinyblock, right=10mm of mulR] (specR) {\LARGE $\hat{s}_{\mathrm{R}}$};
\node[blockISTFT, fill=istftcol, right=10mm of specR] (istftR) {\large ISTFT};

\node[draw, rounded corners=2pt, minimum width=16mm, minimum height=10mm, align=center]
  (outR) at (outX |- istftR.center) {\large Right\\ \large out};

\draw[arrow] (split) |- (rtfR.west);          
\draw[arrow] (rtfR) -- (attnR);               
\draw[arrow] (attnR) -- (unetR);              


\draw[arrow] (split) |- ++(0,-24mm) -| (attnR.south);
\draw[arrow] (unetR) -- node[above] {\LARGE $\mathbf w_{\mathrm R}$} (mulR);

\draw[arrow] (split) |- ++(0,-24mm) -| (mulR.south);

\draw[arrow] (mulR) -- (specR);
\draw[arrow] (specR) -- (istftR);
\draw[arrow] (istftR) -- (outR);

\coordinate (braceX) at ($(outL.east)+(3mm,0)$);
\coordinate (braceTop) at (braceX |- outL.north);
\coordinate (braceBot) at (braceX |- outR.south);

\draw[decorate, decoration={brace, amplitude=6pt}, thick]
  (braceTop) -- (braceBot)
  node[midway, right=4mm, align=center]
  {\large \textbf{Binaural}\\ \large \textbf{output}};

\node[align=center] at ($(unetL.north)!0.5!(rtfL.north)+(0,14mm)$) {\large  \textbf{ Left branch}};
\node[align=center] at ($(unetR.south)!0.5!(rtfR.south)+(0,-14mm)$) {\large \textbf{ Right branch}};

\end{tikzpicture}
\end{adjustbox}
\setlength{\belowcaptionskip}{-12pt}
\setlength{\abovecaptionskip}{-10pt}
\caption{Overview of the proposed dual-branch beamforming network.}
\label{fig:proposed model}
\end{figure*}
\subsection{Time-varying RTF estimation}

To estimate the time-varying RTF of the speaker, we employed a recursive estimation procedure based on the projection approximation subspace tracking (PAST) algorithm \cite{Yang:PAST} and the covariance-whitening (CW) method for RTF estimation \cite{Markovich:Covariance_Whitening_Analysis}.

\subsubsection{Covariance-Whitening}

The CW approach first estimates the noise covariance matrix from $L_n$ noise-only frames (assumed to be available):
\begin{equation}\label{eq: Noise Correlation Estimation_short}
\hat{\Mat{\Phi}}_{\vct{nn}}(k) = \frac{1}{L_n} \sum_{l=0}^{L_n-1}\vct{y}(l,k)\vct{y}^\rmH(l,k).
\end{equation}
The noisy signal covariance matrix is then estimated as:
\begin{equation}\label{eq: Mixed Signal Correlation Estimation_short}
\hat{\Mat{\Phi}}_{\vct{yy}}(k) = \frac{1}{L-L_n}\sum_{l=L_n}^{L-1}\vct{y}(l,k)\vct{y}^\rmH(l,k).
\end{equation}
The whitened measurements are obtained via:
\begin{equation}\label{eq: Whitened Measurement_short}
\vct{y}_{w}(l,k) = \hat{\Mat{\Phi}}^{-1/2}_{\vct{nn}}(k)\,\vct{y}(l,k),
\end{equation}
with $\hat{\Mat{\Phi}}^{-1/2}_{\vct{nn}}(k)$ computed from the eigenvalue decomposition (EVD) of $\hat{\Mat{\Phi}}_{\vct{nn}}(k)$.  
The correlation matrix of the whitened signals is given by:
\begin{equation}
\hat{\Mat{\Phi}}_{\vct{y_{w}y_{w}}}(k) = \hat{\Mat{\Phi}}^{-1/2}_{\vct{nn}}(k)\,
\hat{\Mat{\Phi}}_{\vct{yy}}(k)\,\hat{\Mat{\Phi}}^{-1/2\rmH}_{\vct{nn}}(k).
\end{equation}
Finally, the CW estimate of the RTF is obtained from the principal eigenvector of $\hat{\Mat{\Phi}}_{\vct{y_{w}y_{w}}}(k)$, $\hgvct{\psi}$:
\begin{equation}\label{eq: RTF CW short}
\hvct{a}_{\, \textrm{CW}} \triangleq
\frac{\hat{\Mat{\Phi}}^{\rmH/2}_{\vct{nn}}(k)\hgvct{\psi}}
{\vct{e}^{T}_{{\mathrm{ref}}}\hat{\Mat{\Phi}}^{\rmH/2}_{\vct{nn}}(k)\hgvct{\psi}},
\end{equation}
where $\vct{e}_{{\mathrm{ref}}}$ is the selection vector for the chosen reference microphone. 
The application of the EVD requires $\mathcal{O}(M^3)$ operations per frequency, in addition to the computational cost of the whitening procedure.

\subsubsection{Projection Approximation Subspace Tracking}


\begin{algorithm}[h]
\caption{PAST Algorithm for Tracking the Principal Eigenvector. \cite{Yang:PAST}}
\label{alg:past}
\begin{algorithmic}[1]
\STATE Initialize $\delta(0)$ and $\boldsymbol{\psi}(0)$ 
\FOR{each time-frame $l = 1, 2, \dots$}
    \STATE $\alpha(l,k) = \boldsymbol{\psi}^{\rmH}(l-1,k)\,\mathbf{y}_w(l,k)$
    \STATE $\delta(l,k) = \beta \,\delta(l-1,k) + |\alpha(l,k)|^2$
    \STATE $\mathbf{e}(l,k) = \mathbf{y}_w(l,k) - \boldsymbol{\psi}(l-1,k)\,\alpha(l,k)$
    \STATE $\boldsymbol{\psi}(l,k) = \boldsymbol{\psi}(l-1,k) 
        + \tfrac{\mathbf{e}(l,k)\,\alpha^{*}(l,k)}{\delta(l,k)}$
\ENDFOR
\end{algorithmic}
\end{algorithm}
Since the (whitened) signal is non-stationary, a tracking algorithm is required to estimate the time-varying principal eigenvector $\hgvct{\psi}(l,k)$. The PAST algorithm \cite{Yang:PAST} offers an efficient recursive estimator of the dominant eigenvector of a time-varying correlation matrix. Unlike batch EVD, it updates the subspace incrementally as new data arrive, making it well-suited to real-time causal speech enhancement and beamforming in dynamically evolving acoustic scenes. The application of the PAST algorithm only requires $\mathcal{O}(M)$ operations per frequency (in addition to the whitening procedure). 
 A forgetting factor $\beta \in (0,1)$ is used to gradually discard older observations, allowing the algorithm to adapt to changes in the signal's statistics. 
The PAST procedure is summarized in Algorithm~\ref{alg:past}.

%


\section{Experimental Study}
\label{sec:experiments}


\begin{figure*}[t]
    \centering
    \begin{minipage}{0.98\textwidth}
        \centering
        \includegraphics[width=0.333\textwidth]{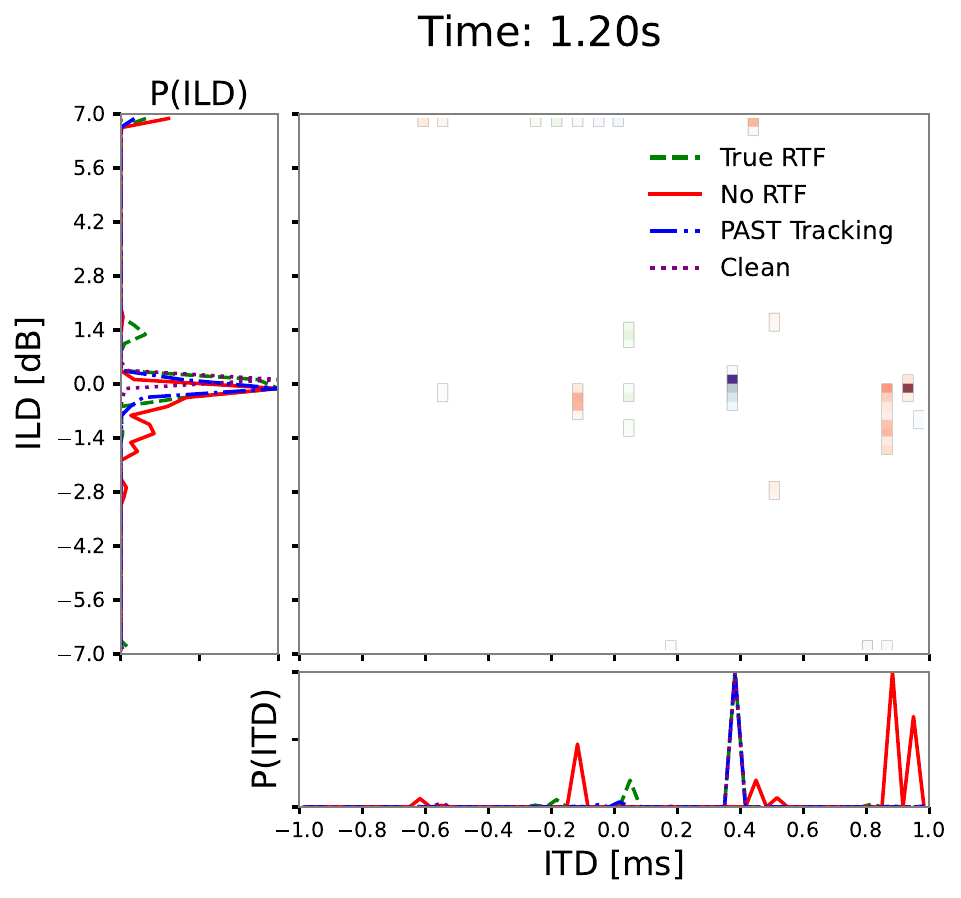}%
        \includegraphics[width=0.333\textwidth]{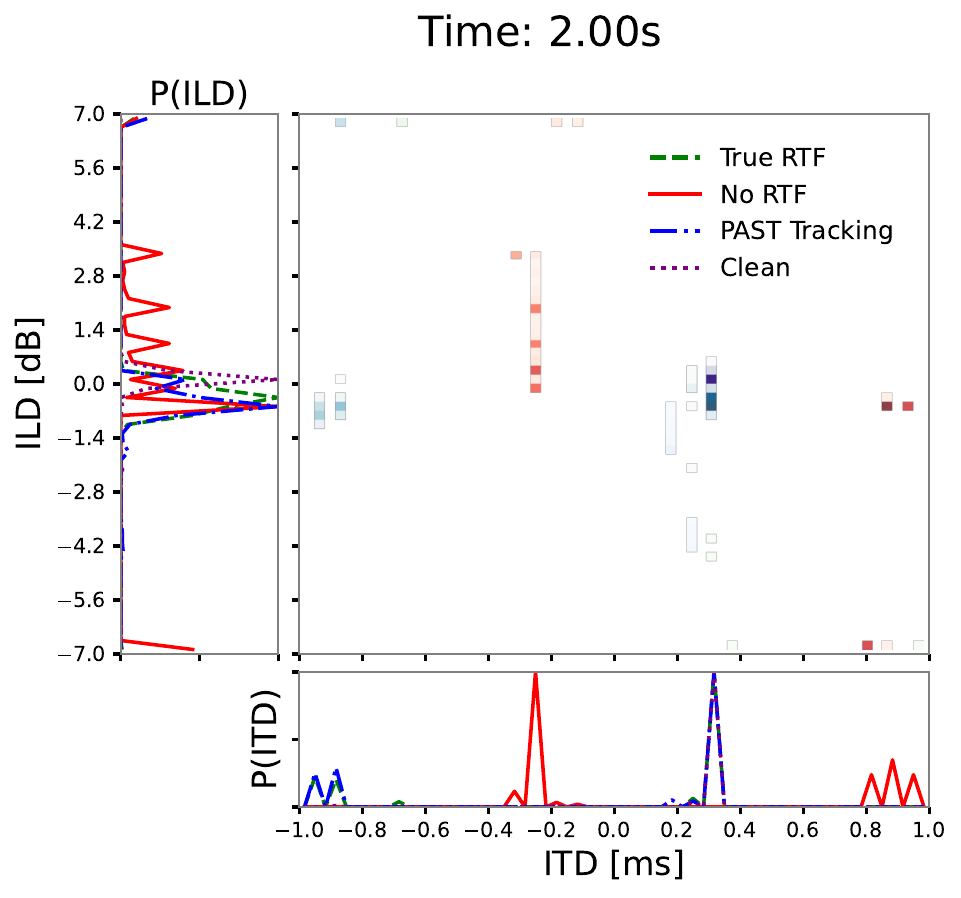}%
        \includegraphics[width=0.333\textwidth]{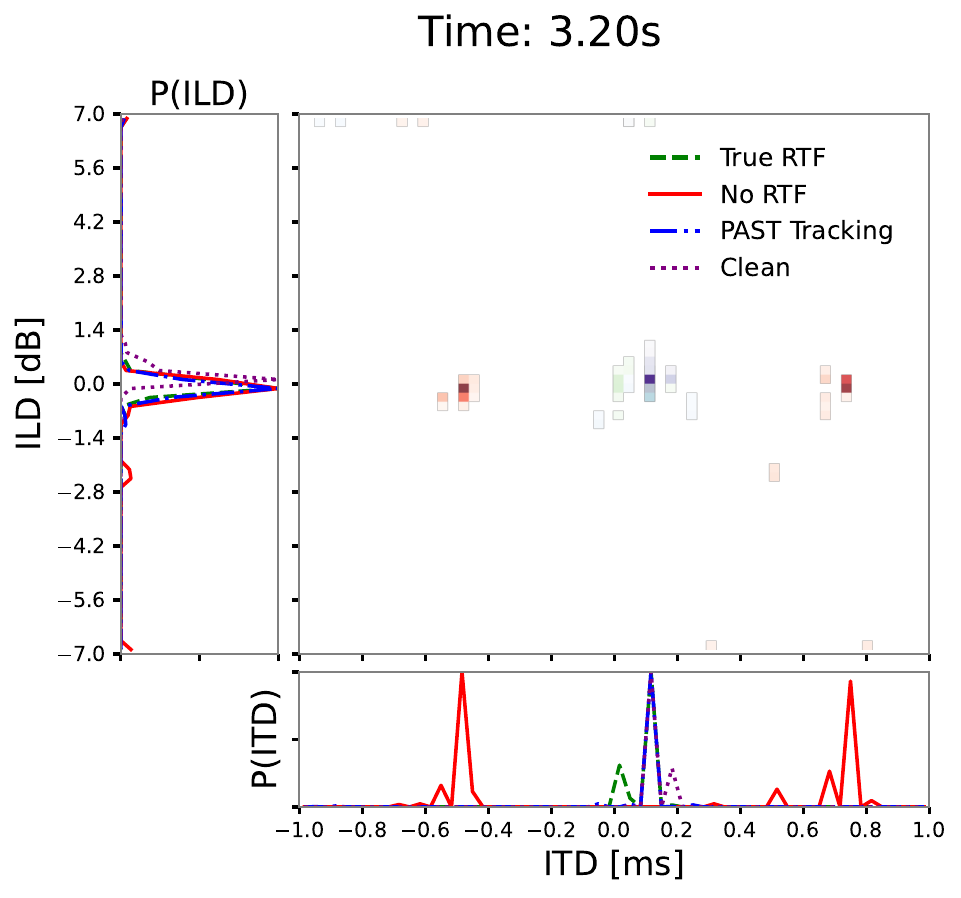}

        \caption{ILD/ITD graphs comparing the three model variants. Graphs produced by \cite{faller2004source} (center frequency $f_{\rm c}=500~\mathrm{Hz}$).}
        \label{fig:joint_3model}
    \end{minipage}
\end{figure*}
In this section, we describe the data generation process, the experimental setup,
and the results of the proposed model.
\subsection{8-channel array data generation}
The dataset was generated with the \textit{Signal Generator} package to simulate moving speakers.\footnote{\texttt{https://github.com/ehabets/Signal-Generator}}
Each sample corresponds to a low-reverberant room with width/length uniformly drawn in $[6,9]$~m and fixed height of $3$~m. An $8$-microphone linear array was placed at height $1.3$~m and randomly tilted within $[-45^\circ,45^\circ]$ (see Fig.~3 in \cite{Cohen:ExNet_BF_PF} for the array configuration).
Target speech was taken from the LibriSpeech dataset \cite{panayotov2015librispeech} and placed $1$–$1.5$~m from the array center. The source moved along a circular trajectory, with its azimuth sweeping linearly from a random initial angle $\theta_0$ to $\theta_0 + \Delta\theta$, where $\Delta\theta \sim \mathcal{U}\big(\pm[45^\circ,150^\circ]\big)$.
Babble noise was simulated by summing $20$ simultaneously active speakers positioned near the room walls; each babbler was filtered using the room impulse response (RIR) generator \cite{habets2006room}.
Overall, the training set contains $20{,}000$ multichannel recordings of duration $4.5$\,s.

\subsection{Binaural dynamic speaker with HRTF simulator}

To simulate a dynamic target speaker with HRTF-based binaural acoustics, we modified the \textit{Signal Generator} to operate in an anechoic binaural setting using HRTFs stored in the Spatially Oriented Format for Acoustics (SOFA) \cite{SOFA_paper}. The listener is modeled as a fixed head at the origin, and the binaural signals correspond to the left and right ears defined by the selected HRTF database; throughout this work, we use the RIEC HRTF dataset \cite{RIEC_HRTF_Website}.

For each utterance, a subject-specific SOFA file is selected, providing head-related impulse responses (HRIRs) on a discrete azimuth-elevation grid. The target speaker moves along a circular trajectory at a fixed radius, with azimuth varying linearly between start and end angles while elevation remains constant. At each update time (once per STFT frame), the instantaneous source direction is computed and the nearest HRIR on the SOFA grid is selected. The binaural signals are generated by convolving the clean speech with this time-varying, piecewise-constant HRIR sequence, yielding an efficient dynamic binaural simulation that preserves spatiotemporal cues in an anechoic environment.
This differs from the original \textit{Signal Generator}, which filters signals using time-varying multi-microphone RIRs computed via the image-source method, including reflections and late reverberation.
\subsection{Binaural reverberant babble noise environment}
In addition to the binaural dynamic target speaker, spatially diffused babble noise is generated using the \textit{SofaMyRoom} simulator \cite{sofamyroom}, which produces echoic binaural RIRs based on SOFA-formatted HRTFs, including early reflections and reverberation.
For each scene, the same subject-specific SOFA file is used for both target and babble signals to ensure consistent head-related spatial cues.
The room parameters match those of the 8-channel case.

\subsection{Loss Function}
For each ear, we maximize the SI-SDR between the beamformer output $\hat{s}_{c}$ and the clean reference signal $s_{\mathrm{ref},c}$ received at the corresponding reference microphone. Additionally, we penalize for the residual noise power at the output. The parameters $\alpha$ and $\lambda$ are tunable.
%
\begin{equation}  
\label{eq:loss_short}
\mathcal{L}
=\sum_{c\in\{\mathrm{L},\mathrm{R}\}}
\left(
-\alpha\;\mathrm{SI\text{-}SDR}(\hat{s}_{c}, s_{\mathrm{ref},c})
+\lambda\mathbb{E}\!\left[\|\mathbf{w}_c^{\mathrm{H}}\mathbf{n}\|_2^2\right]
\right).
\end{equation}

\subsection{Results}
This section reports the results for three model configurations: (i) oracle guidance with true RTFs, (ii) RTF guidance with estimated RTFs obtained by the PAST algorithm (Alg.~\ref{alg:past}), and (iii) a baseline without RTF input. 
Audio samples are available on our demo page.\footnote{\texttt{https://ilaizaidel.github.io/BinDynBeam/}}
\subsubsection{SI-SDR Test Results}
SI-SDR results are presented in Table~\ref{tab:MODEL_RESULTS}.
For the 8-channel case, the results indicate that incorporating the RTF does not substantially affect SI-SDR performance. The oracle RTF achieves the best results among the three configurations, though the improvement over the other cases remains insignificant. In the binaural case, we observe only a slight improvement over the no-RTF case.

\begin{table}[htbp]
\centering
\caption{SI-SDR Test Results ($\mathrm{dB}$). }
\label{tab:MODEL_RESULTS}
\begin{tabular}{lccc}
\toprule
\textbf{Model Configuration} & $8$-channel & Binaural  \\
\midrule
Input                        &  $6.6$ & $6.2$ \\
\midrule
Proposed, no RTF             & $11.9$ & $9.9$ \\
Proposed w. PAST RTF         &  $11.7$ & $10.3$ \\
\midrule
Proposed, oracle             & $12.0$ & $10.3$ \\
\bottomrule
\end{tabular}
\end{table}

\subsubsection{Beampattern Analysis}
\begin{figure}[tb]
    \centering
    \includegraphics[width=0.75\columnwidth]{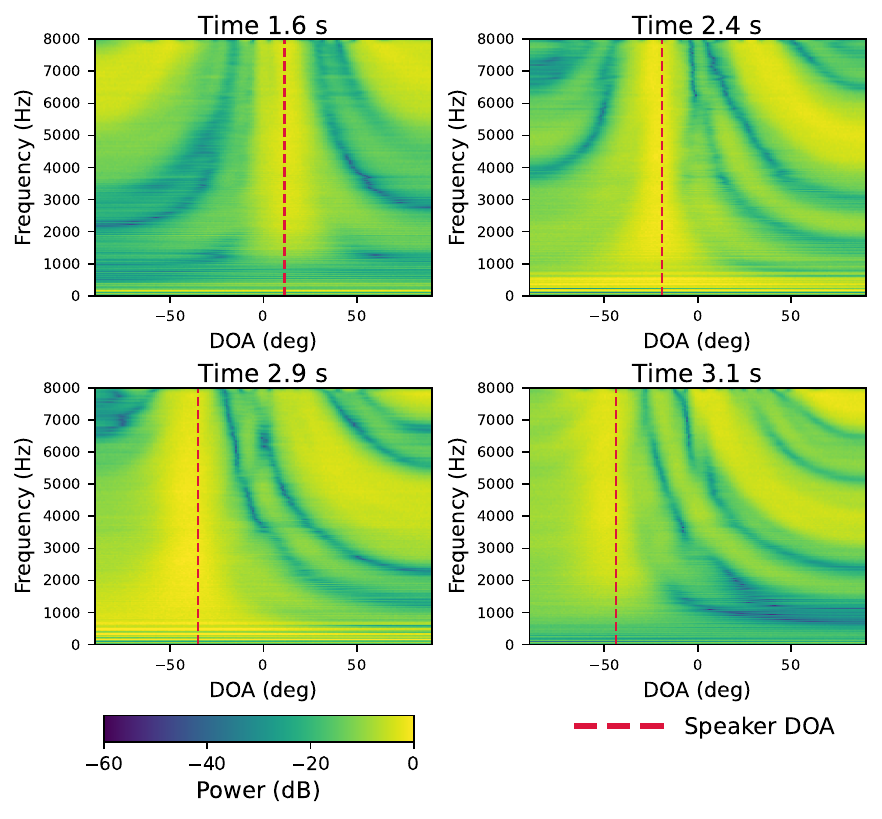}
    \setlength{\belowcaptionskip}{-6pt}
\setlength{\abovecaptionskip}{-2pt}
    \caption{%
    Narrowband time-varying beampattern, at four time snapshots, using RTFs estimated by PAST.
    }
    \label{fig:beampattern_spec_collage_DUAL}
\end{figure}

\begin{figure}[tb]
    \centering
    \includegraphics[width=0.75\columnwidth]{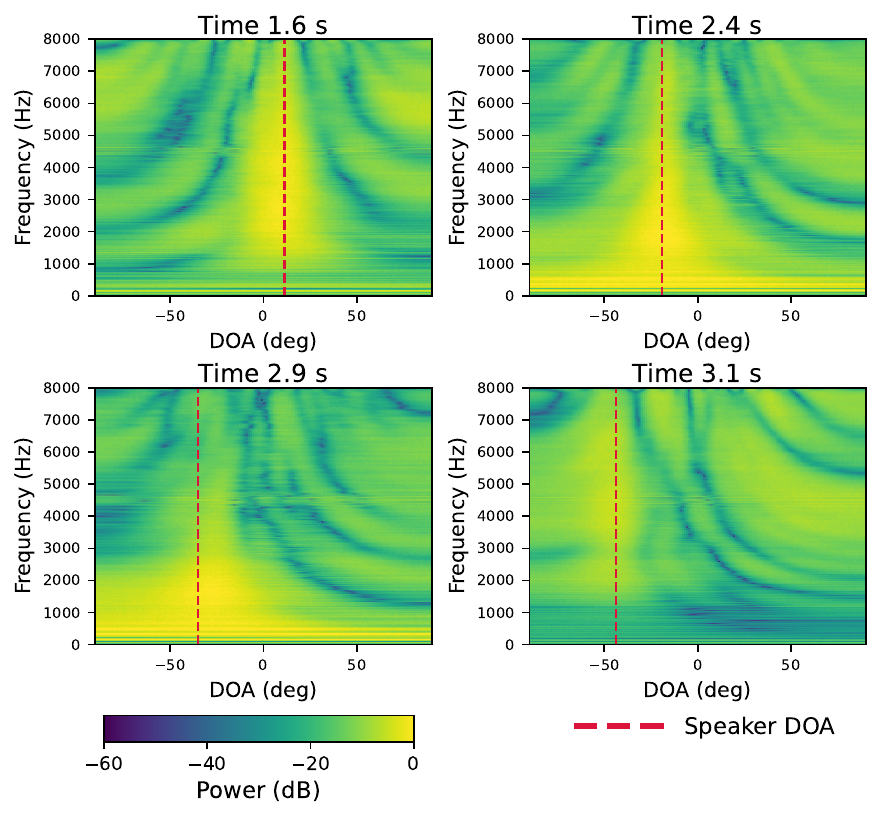}
        \setlength{\belowcaptionskip}{-6pt}
\setlength{\abovecaptionskip}{-2pt}
    \caption{%
    Narrowband time-varying beampattern at four time snapshots, with no RTF guidance.
    }
    \label{fig:beampattern_spec_collage_MIX}
\end{figure}

\begin{figure}[tb]
    \centering
    \includegraphics[width=\linewidth, trim=0 12 5 0, clip]{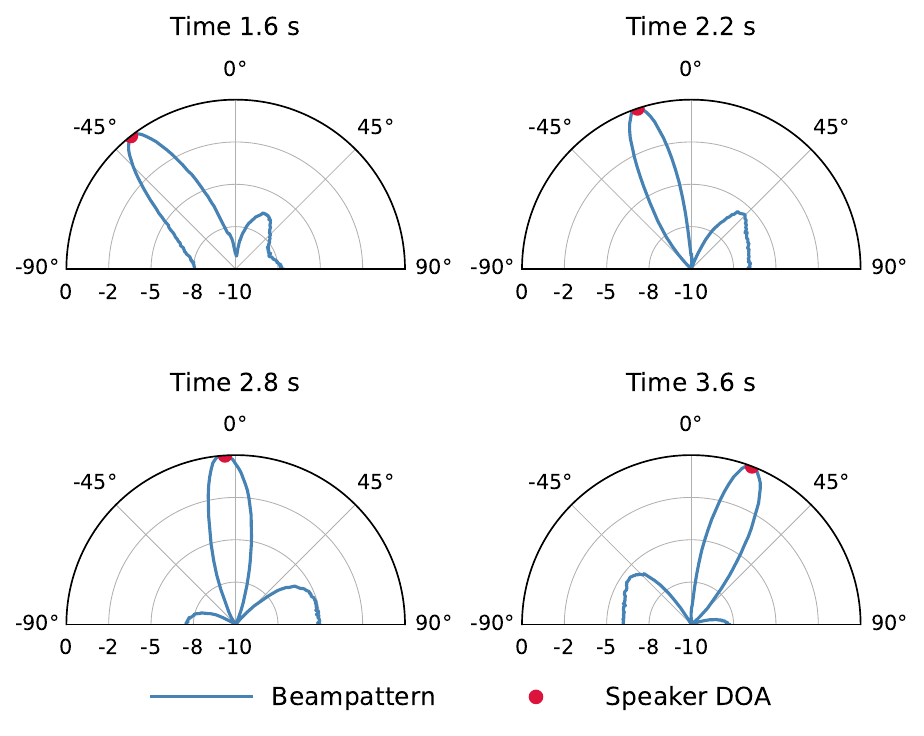}
    \caption{%
    Time-varying wideband beampattern (dB), shown at four time snapshots, with PAST RTF.
    }
    \label{fig:beampattern_collage}
\end{figure}

To further examine the proposed beamformer's spatial behavior, we analyze its beampattern, investigating both narrowband and wideband characteristics.
The time-varying narrowband beampattern is defined as  $B(k,\theta,\ell) = \vct{w}_{\rm L}^{\rm H}(k,\ell)\,\vct{h}(k,\theta),$
where $\vct{h}(k,\theta)$ is the steering vector for DOA $\theta$.  
The corresponding wideband beampower is obtained by a summation over all frequency bins $P(\theta,\ell) = \sum_k \left| B(k,\theta,\ell) \right|^2$.
As shown in Figs.~\ref{fig:beampattern_spec_collage_DUAL},\ref{fig:beampattern_spec_collage_MIX}, guiding the beamformer with the PAST-estimated RTF yields smoother, more spatially coherent beampatterns that tend to follow the moving speaker’s trajectory. In contrast, the model without RTF guidance exhibits reduced directionality and less consistent adaptation to the dynamic scene.
This is further illustrated in Fig.~\ref{fig:beampattern_collage}, which depicts the evolution of the wideband beampattern produced by the PAST RTF-guided model. The main lobe of the beampattern tracks the speaker’s time-varying DOA, indicating that the model adapts to the target’s motion over time.
The static-speaker scenario was compared to the MVDR beamformer in Figs.~5-6 of \cite{Cohen:ExNet_BF_PF} and will not be discussed here.

%




\subsubsection{Binaural cue preservation}

In this section, we analyze the model’s ability to preserve the binaural cues of the dynamically enhanced speaker. To this end, we generate the ILD/ITD curves and compare binaural cue preservation across the three algorithmic variants with respect to the clean reference signal.
Since the speaker is moving, these cues are inherently time-varying. Therefore, the binaural parameters are evaluated over short time segments. Figure~\ref{fig:joint_3model} presents the ILD/ITD trajectories at three representative snapshots along the 4.5~s utterance. The cues are computed within $200~\mathrm{ms}$ analysis windows.
The results demonstrate that incorporating time-varying RTF estimates, either tracked or oracle, substantially improves ITD and ILD accuracy relative to the baseline model without RTF guidance.


\section{Conclusions}
\label{sec:conclusion}


In this paper, we presented an interpretable deep, time-varying beamforming framework for enhancing a moving target speaker in noisy multichannel setups, guided by continuously tracked time-varying RTFs. We evaluated three operating modes: oracle RTF guidance, PAST-based RTF guidance, and no RTF guidance. While SI-SDR remains largely comparable across configurations, RTF guidance consistently improves spatial behavior by yielding smoother, more coherent, and more directional beampatterns that better track the target DOA. We extended the approach to binaural dynamic beamforming. To support this, we have modified  the moving-speaker simulator to incorporate HRTF-based acoustics. We have demonstrated a significantly improved preservation of spatial cues (ILD/ITD) with RTF guidance. 


\balance
\bibliographystyle{IEEEtran}
\bibliography{DNN_BASED_BEAMFORMER}

\end{document}